\documentclass[
  final,5p,times,twocolumn,
  a4paper,
  number,sort&compress,
]{elsarticle}

\usepackage{amssymb}
\usepackage{amsthm}
\usepackage{lineno}
\usepackage{xcolor}

\def\tsc#1{\csdef{#1}{\textsc{\lowercase{#1}}\xspace}}
\tsc{WGM}
\tsc{QE}

\begin{document}




\begin{frontmatter}

\title{PSF Calibration of DAMPE for gamma-ray Observations}  



%


\author[1]{Kai-Kai Duan}
\author[1]{Zhao-Qiang Shen}

\author[1,2]{Zun-Lei Xu}
\author[1]{Wei Jiang}
\author[1,2]{Xiang Li\corref{cor1}}
\ead{xiangli@pmo.ac.cn}






\affiliation[1]{organization={Key Laboratory of Dark Matter and Space Astronomy, Purple Mountain Observatory, Chinese Academy of Sciences},
            addressline={No. 10 Yuanhua Road}, 
            city={Nanjing},
            postcode={210008}, 
            state={Jiangsu},
            country={China}}

\affiliation[2]{organization={School of Astronomy and Space Science, University of Science and Technology of China},
            addressline={No. 3600 Xueyuan Road}, 
            city={Hefei},
            postcode={230026}, 
            state={Anhui},
            country={China}}

\cortext[cor1]{Corresponding author}



\begin{abstract}
The DArk Matter Particle Explorer (DAMPE) is dedicated to exploring critical scientific domains including the indirect detection of dark matter, cosmic ray physics, and gamma ray astronomy. This study introduces a novel method for calibrating the Point Spread Function (PSF) of DAMPE, specifically designed to enhance the accuracy of gamma-ray observations. By leveraging data from regions near pulsars and bright Active Galactic Nuclei (AGNs), we have refined the PSF calibration process, resulting in an improved angular resolution that closely matches our observational data. This advancement significantly boosts the precision of gamma-ray detection by DAMPE, thereby contributing to its mission objectives in dark matter detection and gamma ray astronomy.

\end{abstract}



\begin{keyword}
DAMPE \sep PSF calibration \sep gamma-ray observations
\end{keyword}

\end{frontmatter}


\section{Introduction}\label{intro}
The DArk Matter Particle Explorer (DAMPE) is a gamma-ray telescope that operates through pair conversion, like instruments such as the Energetic Gamma Ray Experiment Telescope (EGRET)~\citep{EGRET}, Astro-rivelatore Gamma a Immagini LEggero (AGILE)~\citep{AGILE}, and Fermi Gamma-ray Space Telescope's Large Area Telescope (Fermi-LAT)~\citep{Fermi2009, Fermi2021}.
Beyond its gamma-ray capabilities, DAMPE is a proficient cosmic-ray detector, adept at measuring charged cosmic rays over a wide energy spectrum~\citep{2014ChJSS..34..550C, 2017APh....95....6C}.
Its large effective area and superior charge and energy resolution enable precise measurements of cosmic-ray spectra and stringent constraints on exotic particles.~\citep[e.g.][]{DAMPE2017e, DAMPE2019p, DAMPE2021He, DAMPE2021solar, DAMPE2022BC, DAMPE2022fcp, DAMPE2023pHe}.
DAMPE's exceptional energy resolution and high gamma-ray energy threshold allow it to effectively detect monochromatic and sharp spectral features in the GeV to TeV range~\citep{2022SciBu..67..679D, Liu2022, 2023PhRvD.108f3015C,Fan:2024rcr} and accurately measure diffuse gamma-ray {emission} up to TeV energies.
Furthermore, DAMPE can contribute to the study of high-energy transients like Active Galactic Nuclei (AGN), {pulsar} flares, and bright gamma-ray bursts, making it a significant tool in astrophysics~\citep{Lei2017, LiX2019, Duan2023ICRC}.

DAMPE consists of four sub-detectors: the Plastic Scintillator strip Detector (PSD), which identifies charged particles; Silicon-Tungsten tracker-converter (STK), {which converts gamma-rays to electron-positron pairs and tracks them}; BGO imaging calorimeter, which measures the energy of these particles; and Neutron Detector (NUD), which detects neutrons to help to identify {electrons} and hadrons~\citep{2014ChJSS..34..550C, 2017APh....95....6C}.
Gamma-ray photons are converted into electron-positron pairs in tungsten foils, with the tracks of these secondary particles recorded by the detector.
By analyzing these tracks, the original gamma-ray directions can be reconstructed~\citep{2018RAA....18...27X}.
On-orbit calibrations~\citep{2019APh...106...18A, Huang:2020skz,Shen2024} and alignments among sub-detectors~\citep{Andrii2018NIMA, Ma2019, Cui2023} and between the payload and satellite~\citep{Jiang2020RAA} have been completed.

The Point-spread Function (PSF) represents the probability density of the offset between the true and reconstructed gamma-ray photon directions. An accurate understanding of the PSF is essential for modeling gamma-ray sky maps and conducting data analysis in gamma-ray science. The 68\% containment radii, a measure of the PSF's width, are crucial for determining the precision of gamma-ray source localization.

In this study, flight data from bright pulsars and AGNs are analyzed to calibrate DAMPE's PSF. The 68\% containment radii from observational data of these point sources are compared to those from Monte Carlo (MC) simulations. The objective is to verify the MC PSF using flight data and perform calibration if discrepancies are found. The successful calibration of the PSF is vital for enhancing the accuracy of gamma-ray sky maps and improving the detection of high-energy astrophysical phenomena. This work not only validates the performance of DAMPE but also sets the stage for more precise observations and analyses in the future.






\section{PSF fitting with simulation data}\label{simudata}

\begin{figure}[!ht]
  \centering
  \includegraphics[width=0.4\textwidth]{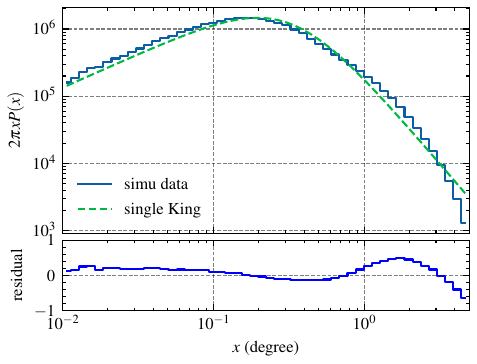}
  \includegraphics[width=0.4\textwidth]{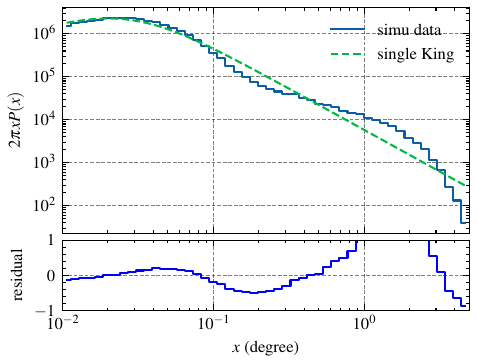}
  \caption{Profiles of PSF distribution fitted with single King function to the simulation data around 2~GeV and 200~GeV {averaged over all angles}.}
  \label{profile_singleKing}
\end{figure}

\begin{figure}[!ht]
  \centering
  \includegraphics[width=0.4\textwidth]{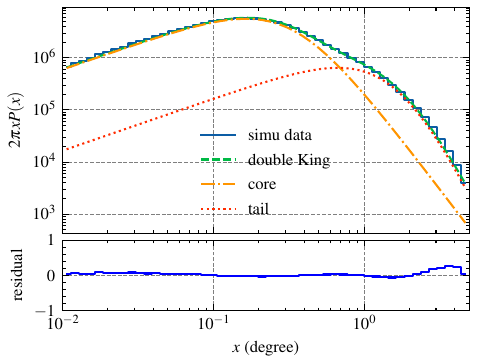}
  \includegraphics[width=0.4\textwidth]{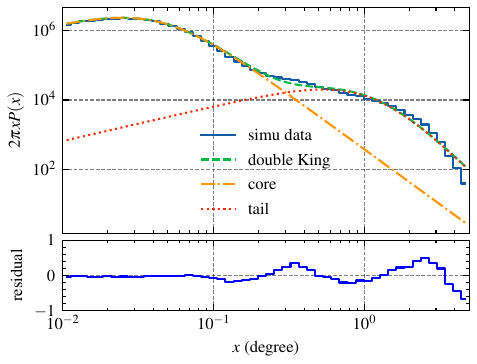}
  \caption{{Profiles of PSF distribution fitted with double King function to the simulation data around 2~GeV and 200~GeV {averaged over all angles}.}}
  \label{profile_doubleKing}
\end{figure}

\begin{figure}[!ht]
  \centering
  \includegraphics[width=0.4\textwidth]{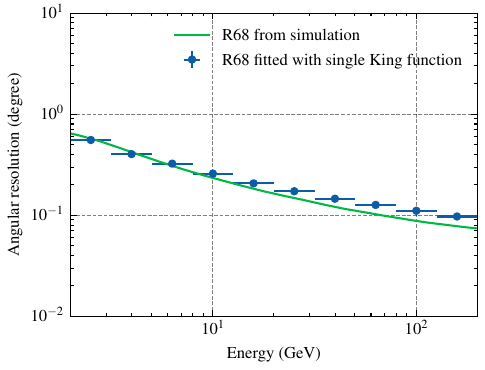}
  \includegraphics[width=0.4\textwidth]{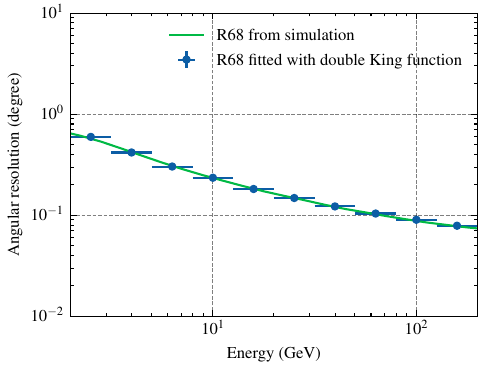}
  \caption{The results of the angular resolution fitted with single ({top}) and double ({bottom}) King function to the simulation data.}
  \label{reuslt_simudata}
\end{figure}

\begin{figure}
    \centering
    \includegraphics[width=0.225\textwidth]{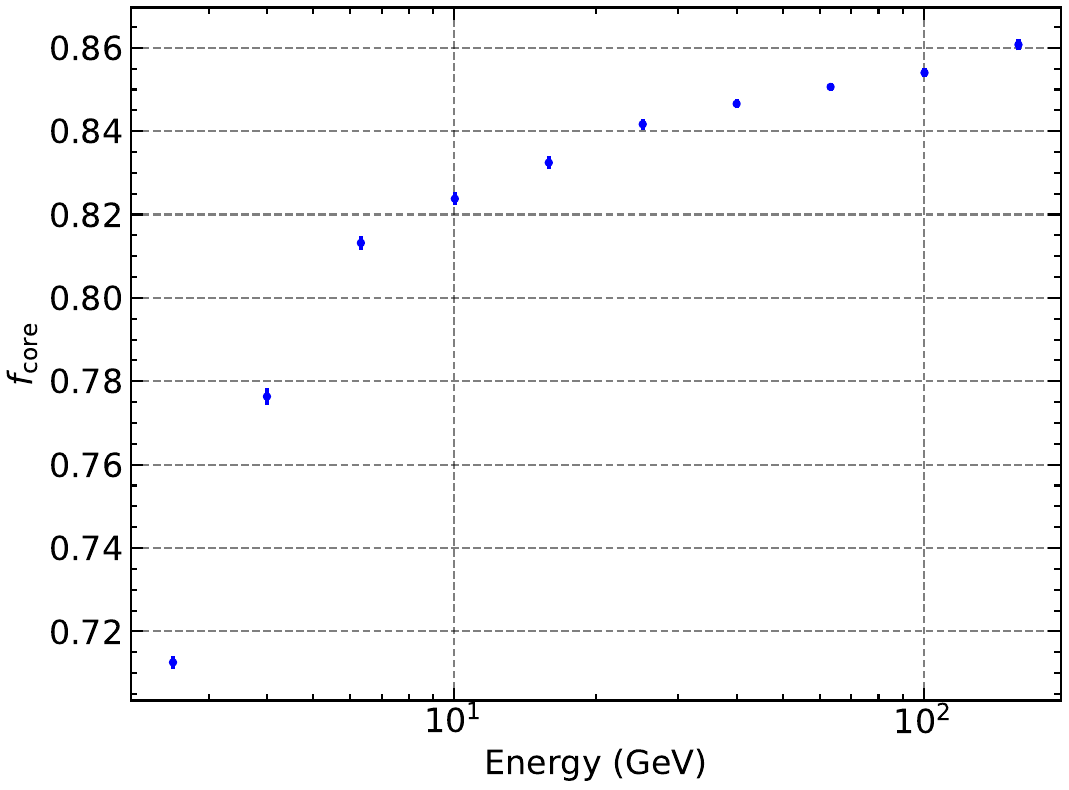}
    \includegraphics[width=0.225\textwidth]{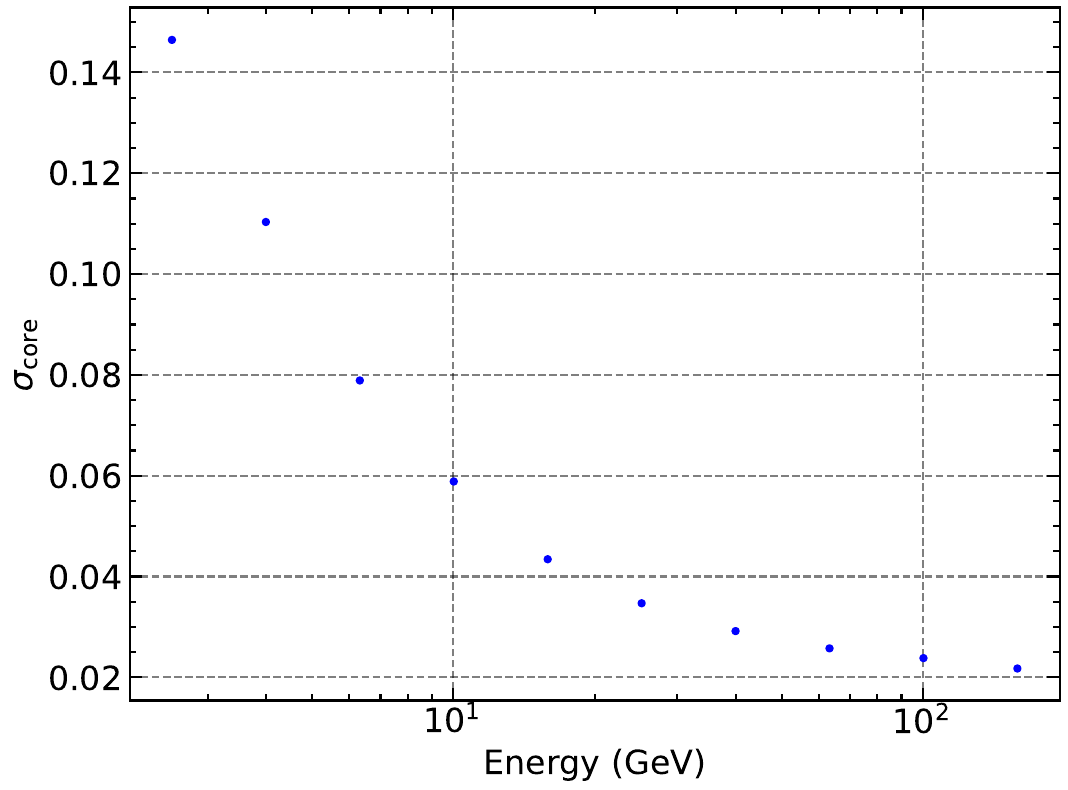}
    \includegraphics[width=0.225\textwidth]{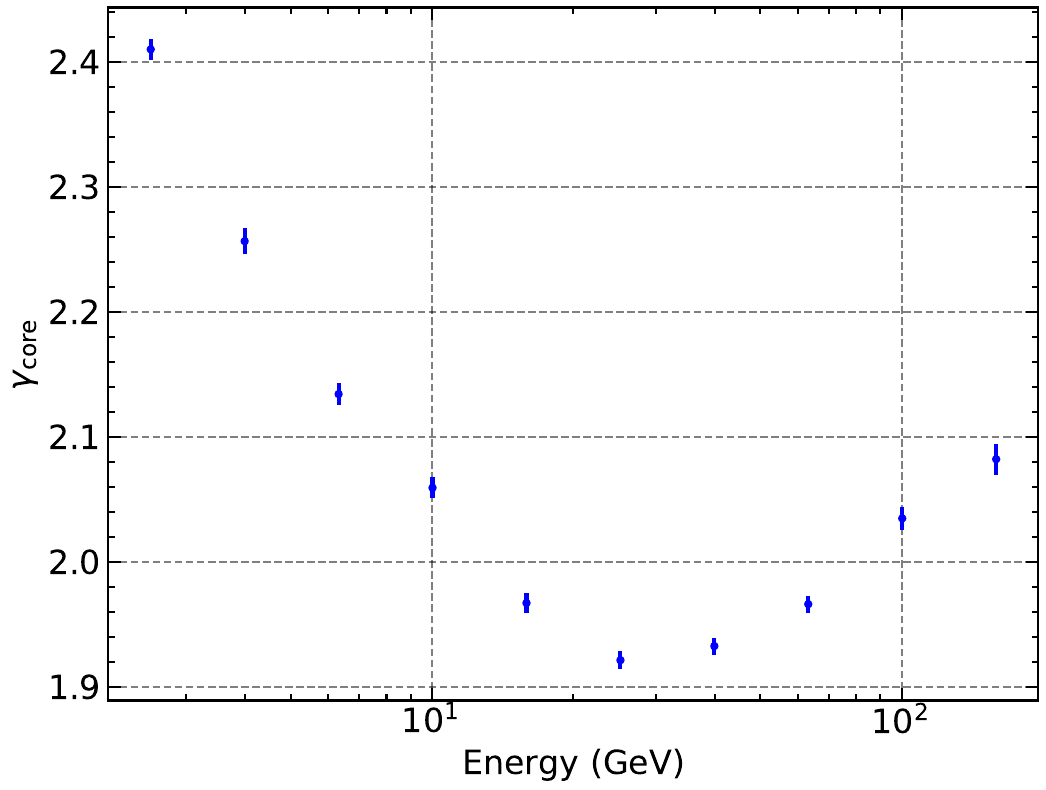}
    \includegraphics[width=0.225\textwidth]{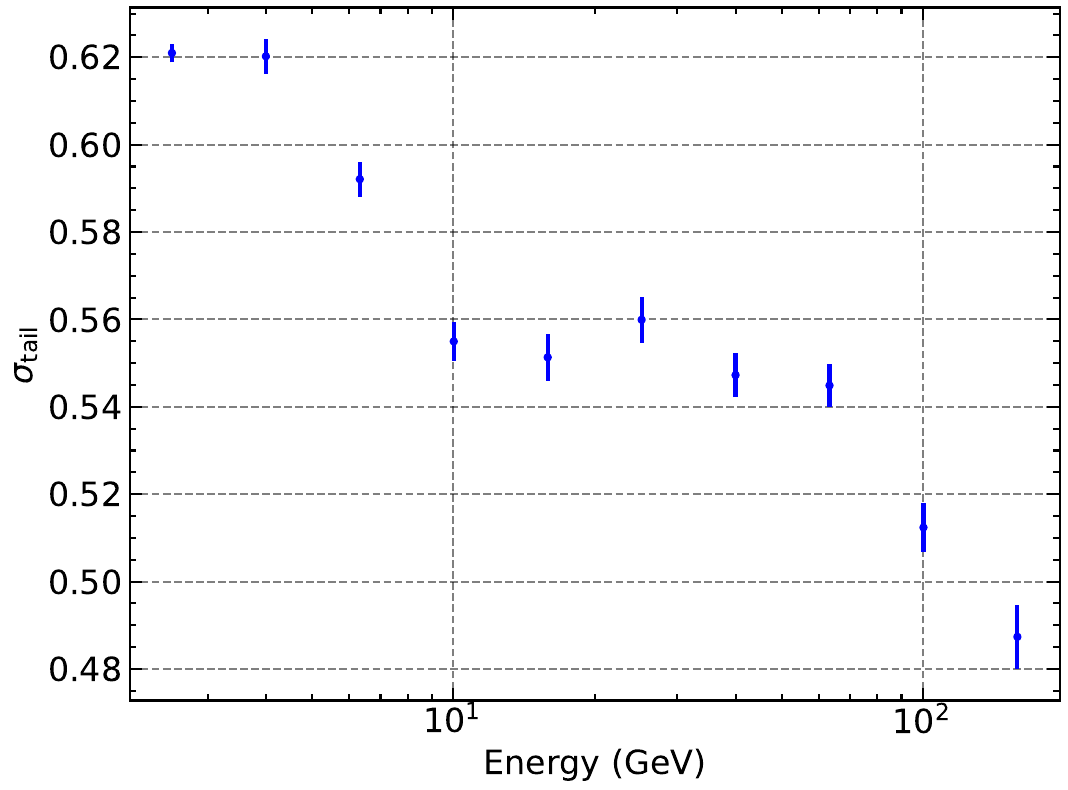}
    \includegraphics[width=0.225\textwidth]{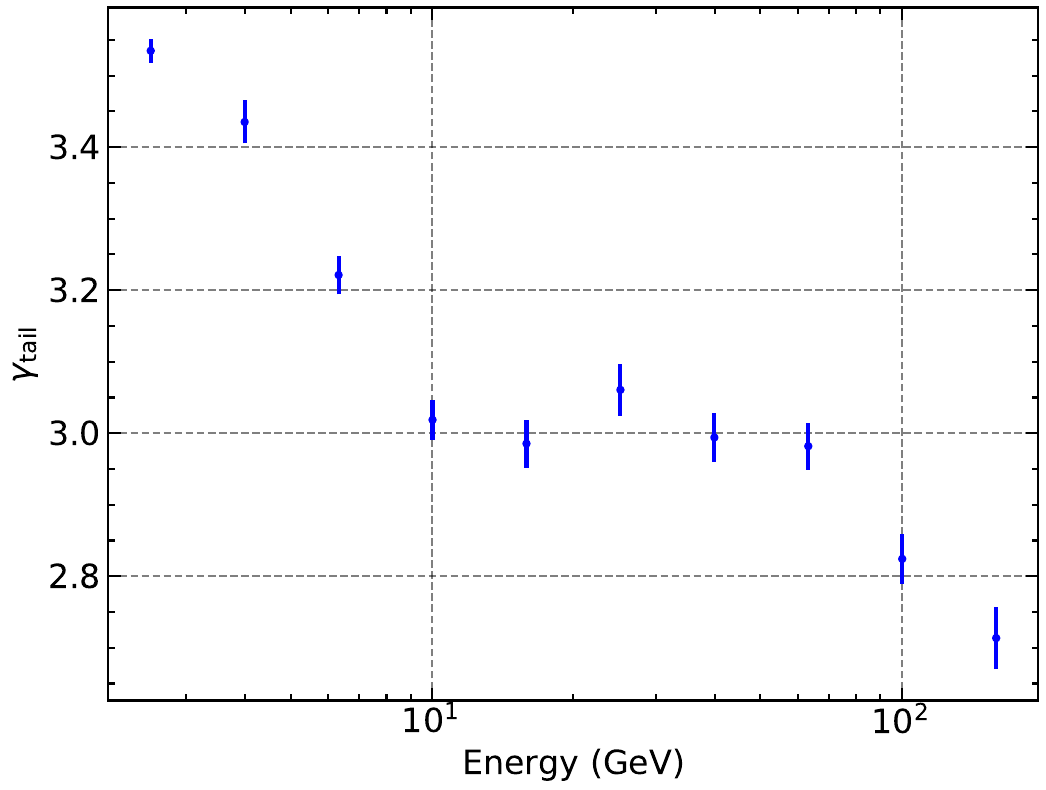}
    \caption{
{The parameters of double King function ($f_{\rm core}$, $\sigma_{\rm core}$, $\gamma_{\rm core}$, $\sigma_{\rm tail}$ and $\gamma_{\rm tail}$) fitted with simulation data.}}
    \label{pars_simudata}
\end{figure}

The PSF distribution in the DAMPE gamma-ray simulation data is parameterized with a double King function~\citep{2019RAA....19..132D}, {which is a commonly used model in gamma-ray astronomy to describe the distribution of detected photons around their source.}
\begin{eqnarray}
    P(x; \sigma_{\mathrm{core}}, \sigma_{\mathrm{tail}}, f_{\mathrm{core}}, \gamma_{\mathrm{core}}, \gamma_{\mathrm{tail}}) = f_{\mathrm{core}} K(x; \sigma_{\mathrm{core}}, \gamma_{\mathrm{core}}) \\
    + (1 - f_{\mathrm{core}}) K(x; \sigma_{\mathrm{tail}}, \gamma_{\mathrm{tail}}),
\end{eqnarray}
where $K(x; \sigma, \gamma)$ is {a} King function~\citep{2004SPIE.5488..103K, 2011A&A...534A..34R} defined as
\begin{eqnarray}
    K(x; \sigma, \gamma) = \frac{1}{2\pi\sigma^{2}} \left( 1-\frac{1}{\gamma} \right) \left[ 1+\frac{1}{2\gamma}\frac{x^{2}}{\sigma^{2}} \right]^{-\gamma},
\end{eqnarray}
and $x$ is the angular deviation {in degrees.} {The $\sigma_{\mathrm{core}}$ and $\sigma_{\mathrm{tail}}$ are also in the unit of degrees.}
Due to limitations in the statistical data from actual flight data, accurately fitting the distribution's tail component presents a significant challenge.
When fitting the simulation data with a single King function, a bias in the angular resolution is observed. 
This bias is illustrated in Figure~\ref{profile_singleKing}, which shows the profiles of the PSF distribution fitted with a single King function.
In contrast, Figure~\ref{profile_doubleKing} displays the PSF profiles fitted with a double King function to the simulation data.
Figure~\ref{reuslt_simudata} explicitly compares the angular resolution obtained from fitting with both single and double King functions, highlighting the non-negligible impact of the tail component in the fitting process.
To address this, in the next section, when fitting the PSF with flight data, the parameters of the tail component will be fixed to align with the results obtained from the simulation data. This approach aims to improve the accuracy of the PSF fitting with flight data by ensuring consistency between simulation and actual data results.
{Figure~\ref{pars_simudata} presents the parameters of double King function ($f_{\rm core}$, $\sigma_{\rm core}$, $\gamma_{\rm core}$, $\sigma_{\rm tail}$ and $\gamma_{\rm tail}$) on energy fitted with simulation data.}

\section{PSF fitting with flight data}
\label{flightdata}

In this research, we calibrate the PSF of the DAMPE instrument using gamma-ray events recorded between January 1, 2016, and August 1, 2023, with energies exceeding 2 GeV.
We focus on the Vela, Geminga, and Crab pulsars due to their compact size, high gamma-ray flux, and pulsed emission characteristics, making them ideal for calibration.

Photons detected within {a} $5^\circ$ radius around each pulsar's coordinates (Vela~\citep{2004AJ....127.3587F}: $\alpha$ = 128.8361$^\circ$, $\delta$ = -45.1764$^\circ$; Geminga~\citep{1998A&A...329L...1C}: $\alpha$ = 98.4756$^\circ$, $\delta$ = 17.7703$^\circ$; Crab~\citep{2020yCat.1350....0G}: $\alpha$ = 83.6331$^\circ$, $\delta$ = 22.0145$^\circ$) are analyzed.

\begin{figure}[!ht]
  \centering
  \includegraphics[width=0.4\textwidth]{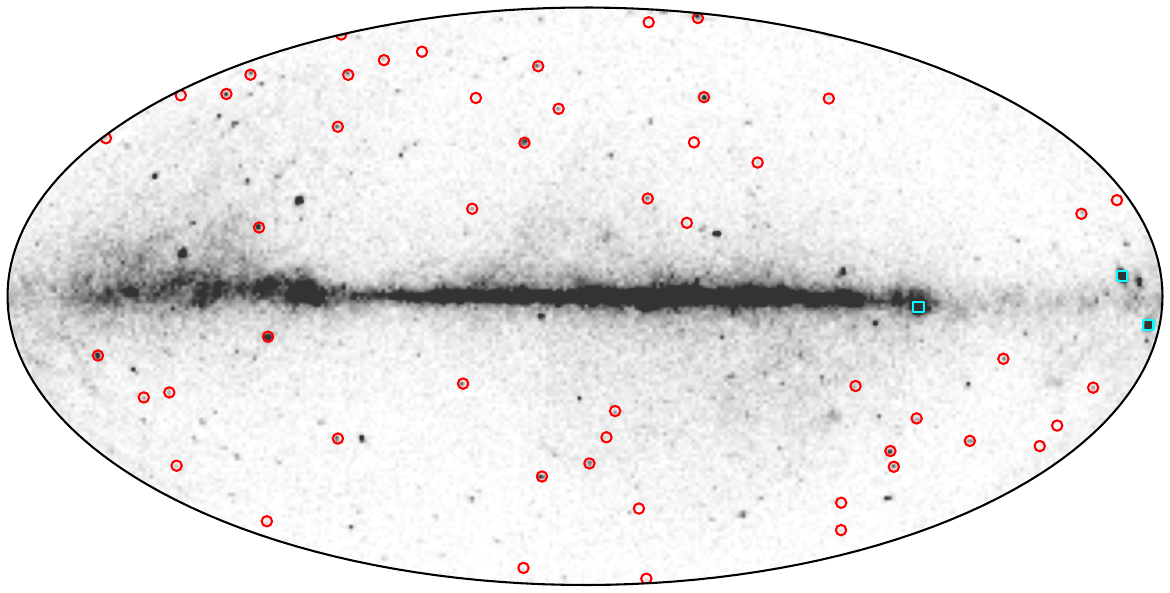}
  \caption{The positions of 51 bright AGNs (red circle) and three pulsars (cyan square) used for PSF calibrations shown in a Hammer-Aitoff projection.}
  \label{fig_agns}
\end{figure}

Additionally, AGNs are utilized for PSF calibration. As the most prevalent gamma-ray sources in the Fermi-LAT 3FHL catalog~\citep{3FHL} and lacking detectable pair-halo emissions~\citep{Ackermann2013ApJ...765...54A}, AGNs are treated as point sources.
We select hard-spectrum AGNs with high-confidence associations from~\citep{Ackermann2012ApJS..203....4A}, excluding those near the Galactic plane to minimize the influence of diffuse Galactic emission.
This selection process yields a sample of 51 AGNs, as shown in Figure~\ref{fig_agns}. The names and galactic coordinates are listed in Table~\ref{tab:my_label}.

\begin{table}[!h]
    \centering
    \begin{tabular}{c|c|c}
    \hline
    Source  & l ($^{\circ}$)    & b ($^{\circ}$) \\
    \hline
    KUV 00311-1938  & 94.17     & -81.21 \\
    PKS 0118-272    & 213.58    & -83.53 \\
    B3 0133+388     & 132.43    & -22.95 \\
    PKS 0215+015    & 162.20    & -54.41 \\
    S4 0218+35      & 142.60    & -23.46 \\
    AO 0235+164     & 156.78    & -39.10 \\
    PKS 0301-243    & 214.64    & -60.17 \\
    NGC 1275        & 150.59    & -13.25 \\
    PMN J0334-3725  & 240.22    & -54.36 \\
    PKS 0420-01     & 195.28    & -33.15 \\
    PKS 0426-380    & 249.70    & -43.62 \\
    PKS 0440-00     & 197.21    & -28.44 \\
    PKS 0447-439    & 248.81    & -39.91 \\
    PKS 0454-234    & 223.73    & -34.90 \\
    TXS 0506+056    & 195.40    & -19.62 \\
    PKS 0537-441    & 250.08    & -31.09 \\
    TXS 0628-240    & 232.68    & -15.00 \\
    PKS 0700-661    & 276.77    & -23.78 \\
    B2 0716+33      & 185.06    & 19.85 \\
    PKS 0735+17     & 201.85    & 18.06 \\
    S4 0814+42      & 178.21    & 33.41 \\
    S4 0917+44      & 175.70    & 44.81 \\
    4C +55.17       & 158.59    & 47.94 \\
    1H 1013+498     & 165.53    & 52.73 \\
    4C +01.28       & 251.50    & 52.77 \\
    Mkn 421         & 179.82    & 65.03 \\
    Ton 599         & 199.41    & 78.37 \\
    4C +21.35       & 255.07    & 81.66 \\
    PKS 1244-255    & 301.60    & 37.08 \\
    PG 1246+586     & 123.74    & 58.77 \\
    S4 1250+53      & 122.36    & 64.08 \\
    3C 279          & 305.10    & 57.06 \\
    GB 1310+487     & 113.32    & 68.25 \\
    PMN J1344-1723  & 320.48    & 43.67 \\
    PKS 1424+240    & 29.48     & 68.20 \\
    PKS 1454-354    & 329.89    & 20.52 \\
    PKS 1502+106    & 11.37     & 54.58 \\
    AP Librae       & 340.70    & 27.58 \\
    B2 1520+31      & 50.18     & 57.02 \\
    GB6 J1542+6129  & 95.38     & 45.40 \\
    PG 1553+113     & 21.92     & 43.95 \\
    1H 1720+117     & 34.11     & 24.47 \\
    1ES 1959+650    & 98.02     & 17.67 \\
    PKS 2005-489    & 350.37    & -32.61 \\
    PKS 2023-07     & 36.89     & -24.39 \\
    PKS 2052-47     & 352.58    & -40.38 \\
    MH 2136-428     & 358.29    & -48.32 \\
    PKS 2155-304    & 17.74     & -52.24 \\
    BL Lacertae     & 92.60     & -10.46 \\
    3C 454.3        & 86.12     & -38.18 \\
    PKS 2326-502    & 332.00    & -62.30 \\
    \hline
    \end{tabular}
    \caption{{The names and galactic coordinates of the AGNs selected for this work.}}
    \label{tab:my_label}
\end{table}

We divide the observed photons into energy bins: 10 bins from 2 GeV to 20 GeV for pulsars and 8 bins from 20 GeV to 200 GeV for AGNs.
{For each bin, we calculate the angular deviations of photons from the center source within 5$^{\circ}$ for stacking.} 
We conduct a joint unbinned likelihood analysis in each energy bin to model the PSF,
assuming a central point source and an isotropic backgroud, i.e.
{
\begin{eqnarray}
   \lambda(x) &=& N_S \times P(x; f_{\mathrm{core}}, \sigma_{\mathrm{core}}, \gamma_{\mathrm{core}}, \sigma_{\mathrm{tail}}, \gamma_{\mathrm{tail}}) \\
   &+& N_{B}/(\pi x_{\rm max}^2).
\label{eqn_AGNSky}
\end{eqnarray}
}
The parameters $f_{\rm core}$, $\gamma_{\rm core}$, $\sigma_{\rm tail}$ and $\gamma_{\rm tail}$ are fixed to those derived from simulation data that closely match the instrument's expected performance.
The unbinned {Poisson} likelihood we used is
\begin{eqnarray}
\ln \mathcal{L}=\sum_{\rm i}^{n}\ln \left (\lambda (x_{\rm i}) \right )-\int_{\Omega_{\rm x,max}}\lambda(x)d\Omega_{\rm x},
\end{eqnarray}
where $n$ is the number of observed photons from these stacked AGNs and pulsars within selected regions.

\begin{figure}[!tb]
    \centering
    \includegraphics[width=0.225\textwidth]{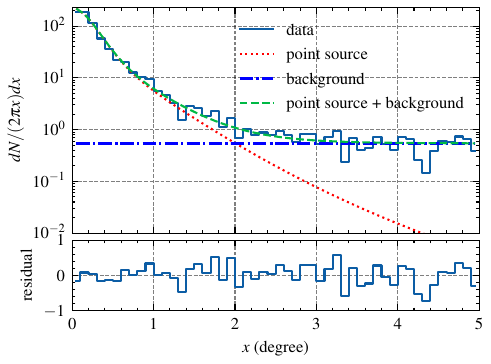}
    \includegraphics[width=0.225\textwidth]{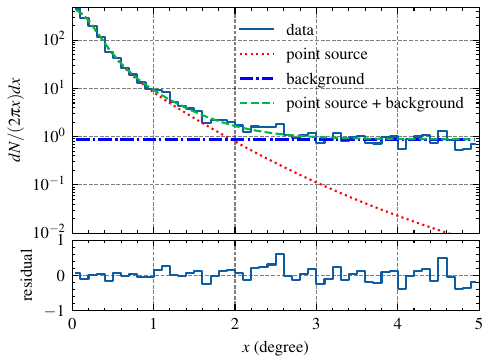}
    \includegraphics[width=0.225\textwidth]{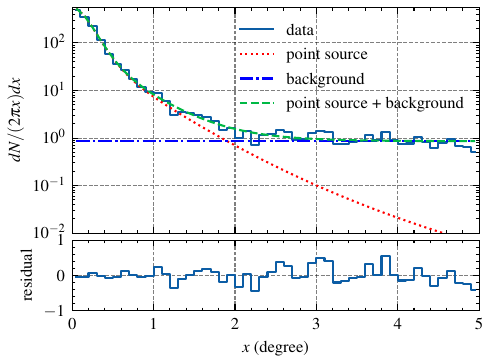}
    \includegraphics[width=0.225\textwidth]{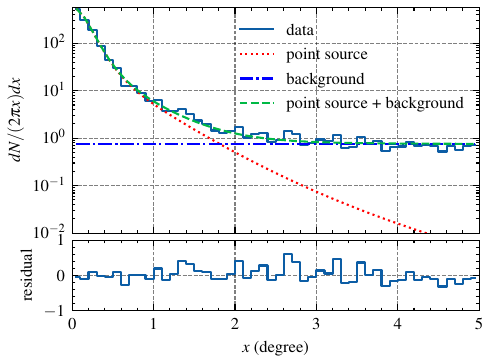}
    \includegraphics[width=0.225\textwidth]{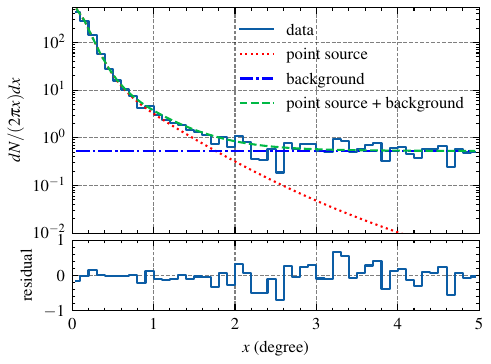}
    \includegraphics[width=0.225\textwidth]{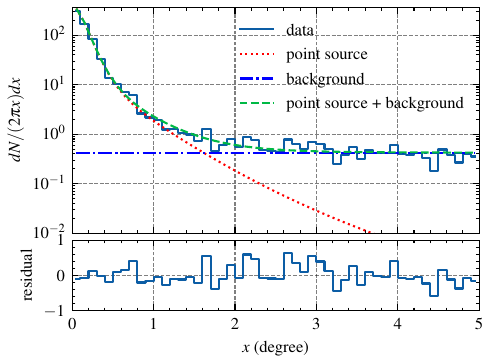}
    \includegraphics[width=0.225\textwidth]{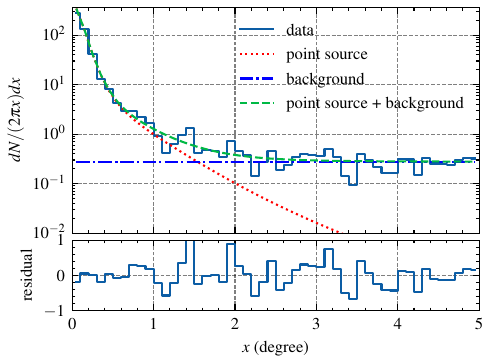}
    \includegraphics[width=0.225\textwidth]{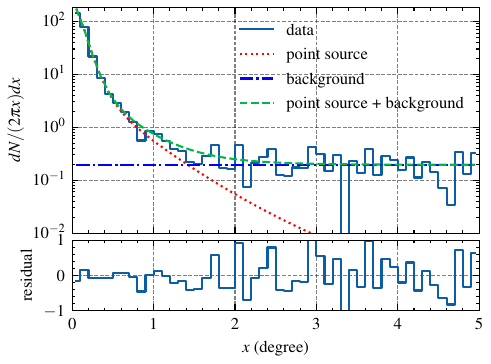}
    \includegraphics[width=0.225\textwidth]{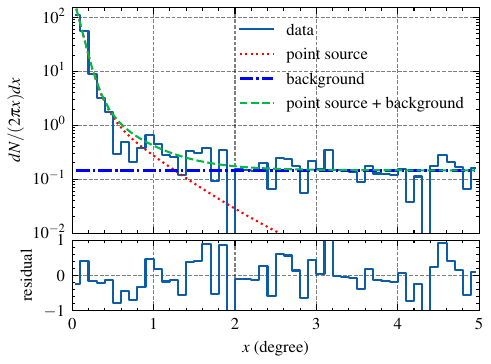}
    \includegraphics[width=0.225\textwidth]{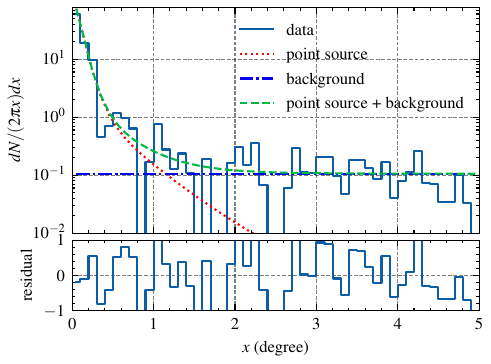}
    \caption{The results of the PSF profiles fitted with data around pulsars in 2.0$-$2.5~GeV, 2.5$-$3.2~GeV, 3.2$-$4.0~GeV, 4.0$-$5.0~GeV, 5.0$-$6.3~GeV, 6.3$-$8.0~GeV, 8.0$-$10.0~GeV, 10.0$-$12.6~GeV, 12.6$-$15.9~GeV and 15.9$-$20.0~GeV {averaged over all angles}. {The residuals are standard deviations.}
    The blue histogram is the gamma-ray photon number density with angular deviation, the red dotted line, blue dashdot line and the blue dashed line are the number densities from central point sources with PSF, the isotropic background and the sum of the point sources and background, respectively.}
    \label{result_doubleKing_Pulsar}
\end{figure}

\begin{figure}[!tb]
    \centering
    \includegraphics[width=0.225\textwidth]{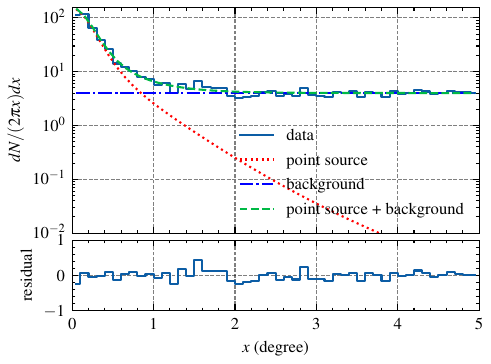}
    \includegraphics[width=0.225\textwidth]{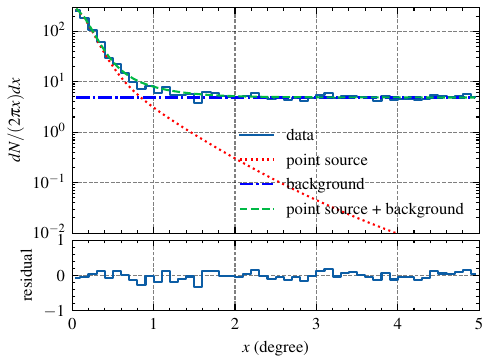}
    \includegraphics[width=0.225\textwidth]{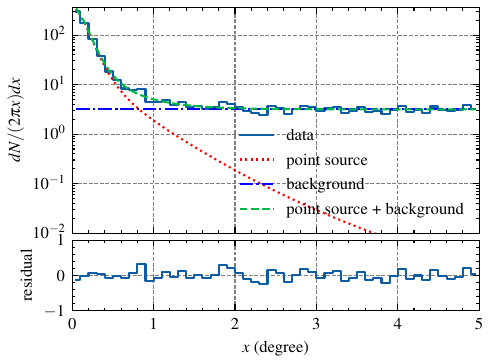}
    \includegraphics[width=0.225\textwidth]{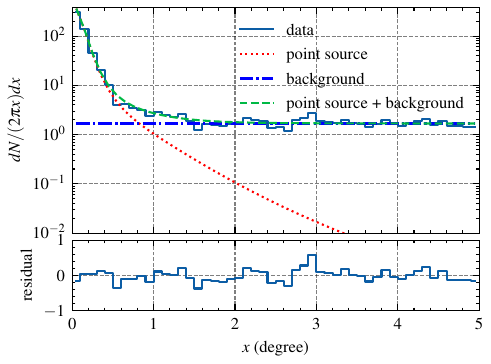}
    \includegraphics[width=0.225\textwidth]{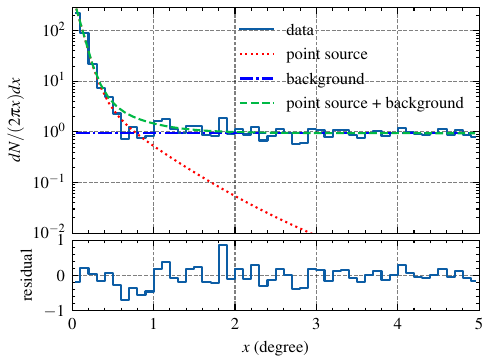}
    \includegraphics[width=0.225\textwidth]{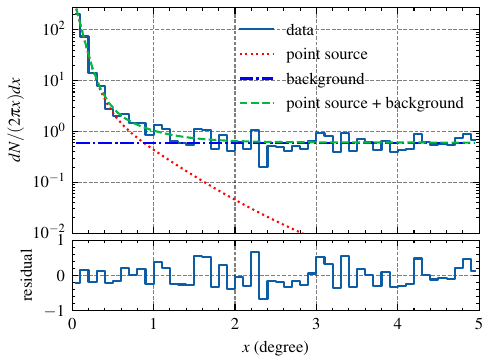}
    \includegraphics[width=0.225\textwidth]{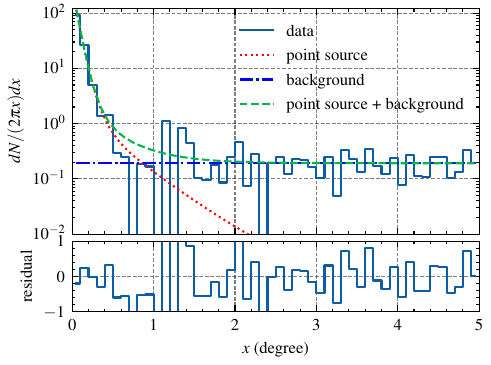}
    \includegraphics[width=0.225\textwidth]{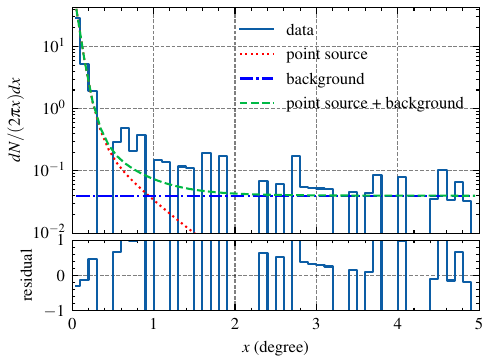}
    \caption{The results of the PSF fitted with data around AGNs from 2.0$-$3.2~GeV, 3.2$-$5.0~GeV, 5.0$-$8.0~GeV, 8.0$-$12.6~GeV, 12.6$-$20.0~GeV, 20.0$-$43.1~GeV, 43.1$-$92.8~GeV and 92.8$-$200.0~GeV {averaged over all angles}. {The residuals are standard deviations.} The legends are the same as Figure~\ref{result_doubleKing_Pulsar}.}
    \label{result_doubleKing_AGN}
\end{figure}

\begin{figure}
    \centering
    \includegraphics[width=0.4\textwidth]{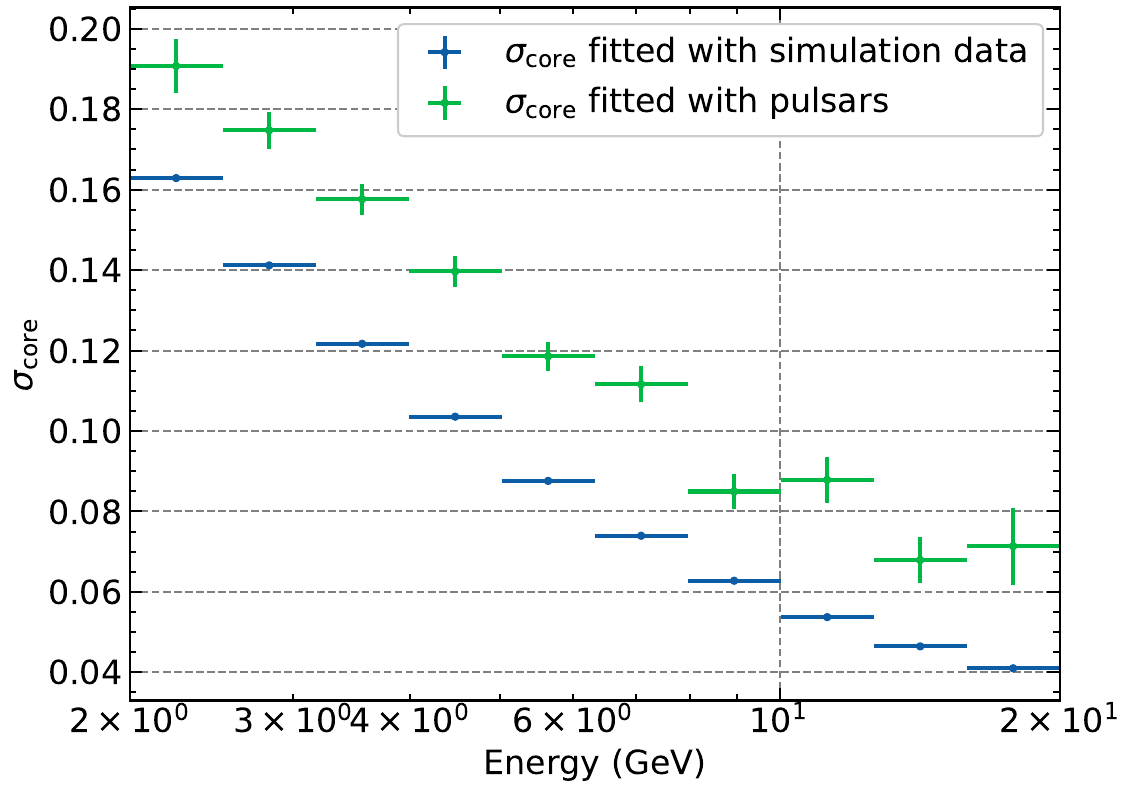}
    \includegraphics[width=0.4\textwidth]{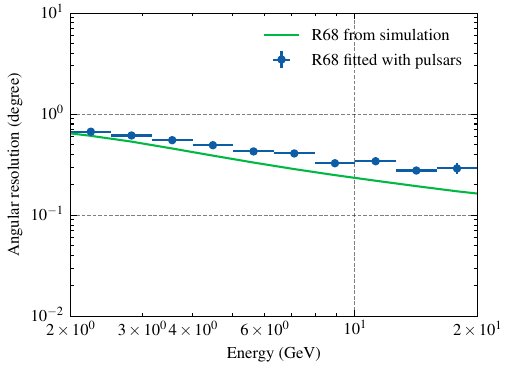}
    \caption{
{The $\sigma_{\rm core}$ parameter (top) and angular resolution (bottom) fitted with pulsars from 2~GeV to 20~GeV compared with simulation.}}
    \label{score_flight}
\end{figure}

\begin{figure}[!ht]
    \centering
    \includegraphics[width=0.38\textwidth]{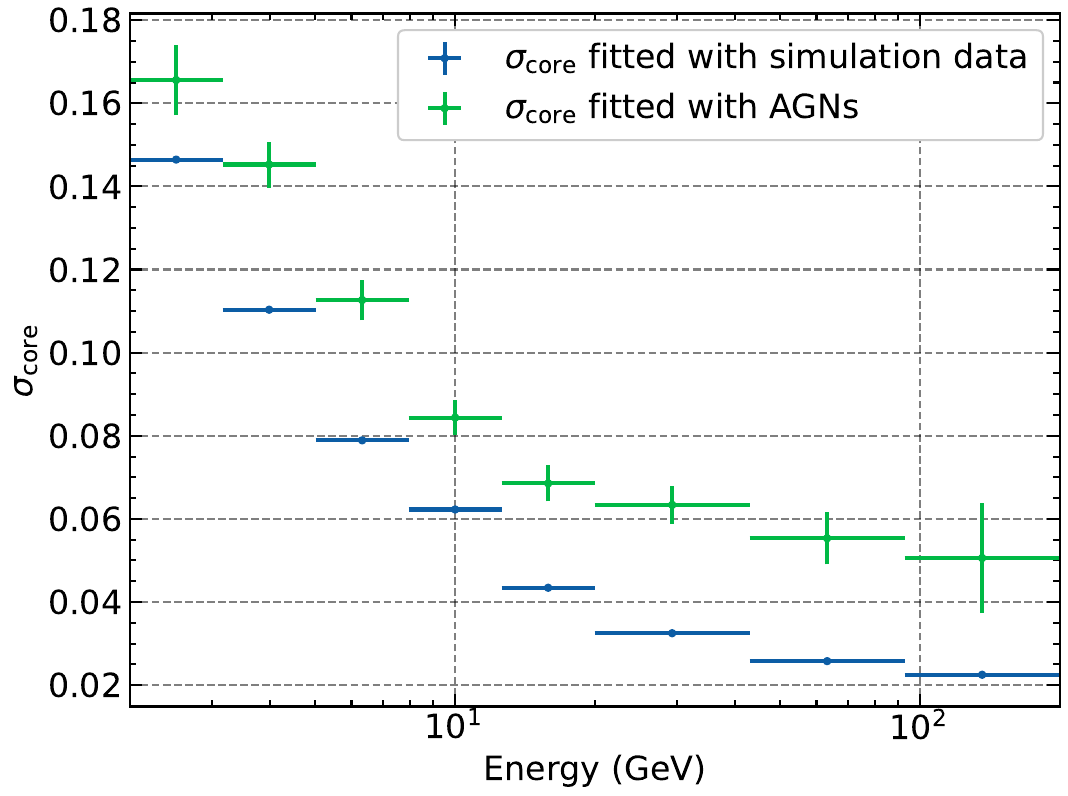}
    \includegraphics[width=0.38\textwidth]{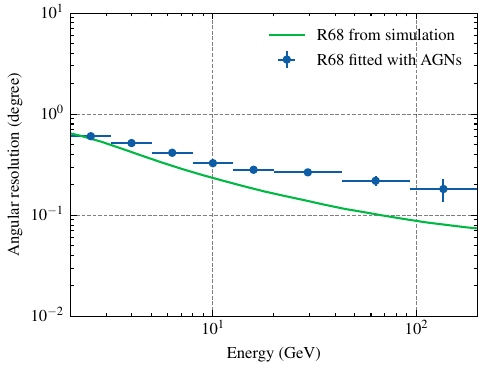}
    \caption{{The $\sigma_{\rm core}$ parameter (top) and angular resolution (bottom) fitted with AGNs from 2~GeV to 200~GeV compared with simulation.}}
    \label{result}
\end{figure}

Figure~\ref{result_doubleKing_Pulsar} and Figure~\ref{result_doubleKing_AGN} illustrate the PSF distributions around pulsars and AGNs, respectively, fitted with a double King function across the specified energy ranges.
{Figure~\ref{score_flight} and Figure~\ref{result} presents the $\sigma_{\rm core}$ parameter and the angular resolution fitted with pulsars and AGNs, respectively.}
{A comparison with simulated angular resolution shows that the PSF in the flight data is broader than predicted by simulations as energy increases.}

\section{Calibration of the PSF with flight data}
\label{calibration}

$\sigma_{\rm core}$, a critical parameter in the double King function used to model the PSF, was adjusted to align the angular resolution predicted by simulations with that observed in flight data. This adjustment was necessary as $\sigma_{\rm core}$ varies with energy, and its scaling is linearly related to the logarithm of energy, as given by 

\begin{equation}
\sigma_{\rm core} / \sigma_{\rm core, MC} =  A + B \times {\log_{10}} (E/{\rm GeV}).
\end{equation}

\begin{figure}
    \centering
    \includegraphics[width=0.45\textwidth]{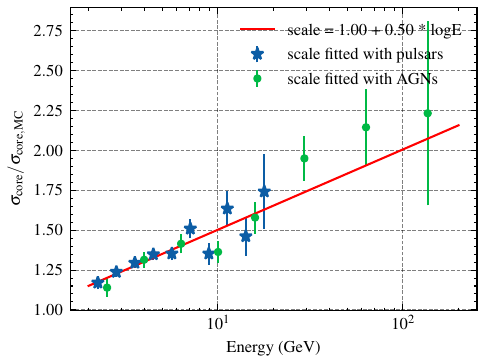}
    \includegraphics[width=0.45\textwidth]{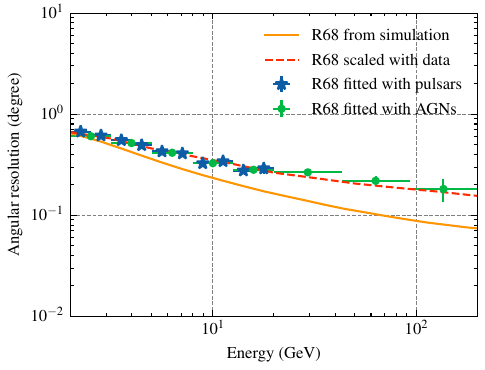}
    \caption{{\it {top panel}:} The scale of $\sigma_{\rm core}$ fitted with the differences between flight data and simulation (red line), the blue stars and green points are the scales fiited with pulsars and AGNs, respectively.
    {\it {bottom panel}:} The result of the angular resolution from the scaled double King function. The blue stars and green points are the angular resolutions fiited with pulsars and AGNs, respectively. The orange solid line and red dashed line are the angular resolutions from simualtion and scaled double King function, respectively.}
    \label{result_scaled}
\end{figure}

\begin{figure}
  \centering
  \includegraphics[width=0.42\textwidth]{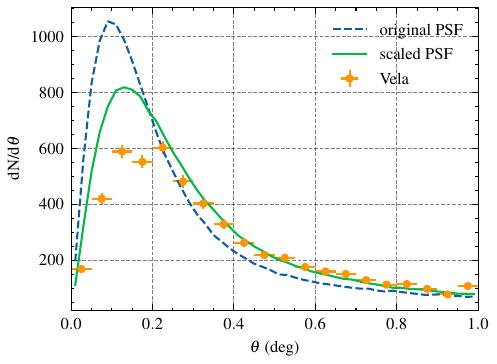}
  \caption{The original and scaled expected photon distributions with the angular distances from the Vela pulsar, compared with the angular distribution of the observed data around Vela pulsar within {a} 1 degree radius. The dashed and solid lines are the original and scaled angular distribution, respectively. The points are the angular distributions of the observed data.}
  \label{Vela}
\end{figure}

Upon analyzing data from pulsars and AGNs, we determined the scaling factors for $\sigma_{\rm core}$ to be $A = 1.00 \pm 0.04$ and $B = 0.50 \pm 0.05$. 
These results are graphically represented in the {top} panel of Figure~\ref{result_scaled}.
The {bottom} panel illustrates the improved angular resolution achieved through the scaling of $\sigma_{\rm core}$ in the double King function.
Our calibration confirms that the angular resolution aligns well with the values obtained from fitting data from pulsars and AGNs.
Figure~\ref{Vela} compares the original and adjusted photon distributions, focusing on the angular distances from the Vela pulsar. The reduced $\chi^2$ values for the original and scaled distributions, when compared to the actual data from the Vela pulsar, are 61 and 10, respectively. These values indicate a significant improvement in fit quality, with a lower chi-squared value suggesting a better fit to the data.

\begin{figure}
    \centering
    \includegraphics[width=0.45\textwidth]{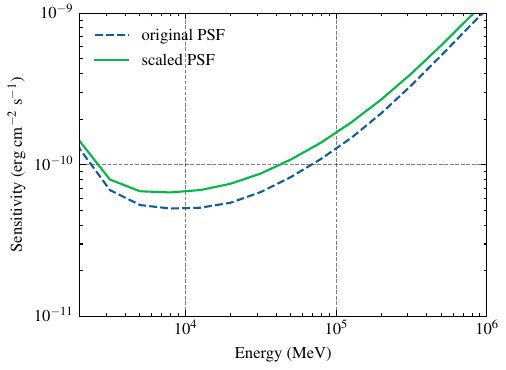}
    \caption{The sensitivity of the gamma-ray data obtained from DAMPE at Galactic Center direction. The blue and green lines are the results obtained with original and scaled PSF, respectively.}
    \label{sensitivity}
\end{figure}

{The sensitivity of the point source is dependent on the effective area of the detector and the background, which is connected with the angular resolution of the detector.
We utilized a point source with an index of $-2$ and diffuse emissions background from the Fermi-LAT results\footnote{https://fermi.gsfc.nasa.gov/ssc/data/access/lat/BackgroundModels.html} to assess the sensitivity at Galactic Center direction. Here we ignore the influence of nearby point sources.}
Figure~\ref{sensitivity} further demonstrates the impact of this calibration on sensitivity, showing that the maximum observed difference in sensitivities between the original and scaled PSF models is 33\% at around 20 GeV. This highlights the importance of accurate PSF calibration for enhancing the precision of gamma-ray observations.

\section{Summary}
\label{summary}

In this research, we evaluate the angular resolution, a measure of the precision with which {DAMPE} can locate the source of a gamma-ray by comparing simulated outcomes with actual data collected near pulsars and bright Active Galactic Nuclei (AGNs).
Our analysis reveals a decline in angular resolution with increasing energy levels, specifically within the range of 2 to 200~GeV.
{The transport process of electron-hole pairs in silicon semiconductors consumes a significant amount of computational time, making it difficult to simulate in detail in large-scale simulations. The usual approach is to parameterize the charge distribution coefficients through detailed semiconductor simulations and to calibrate with ground-based beam tests. This approach may result in slight deviations from the actual situation, leading to discrepancies between trajectory reconstruction and the actual scenario.}
To address this, we propose a scaling method for $\sigma_{\rm core}$ parameter, which is linearly related to the logarithm of energy.
This scaling technique successfully correlates the simulated angular resolution with that derived from in-flight data.
By calibrating the PSF with data from pulsars and AGNs, we confirm that the angular resolution matches the one obtained from flight data, a critical factor for the ongoing analysis of gamma-ray data from DAMPE.
This calibration enhances the reliability of DAMPE's gamma-ray observations, supporting further research and data interpretation in gamma-ray astronomy.

\section*{Acknowledgements}
  We make use of the following Python packages: {\sc NumPy}~\citep{numpy2020}, {\sc SciPy}~\citep{scipy2020}, {\sc Matplotlib}~\citep{matplotlib2007}, {\sc Astropy}~\citep{astropy2018}, {\sc iminuit}~\citep{Minuit1975}, and {\sc DmpST}~\citep{2019RAA....19..132D}.
  The DAMPE mission was funded by the strategic priority science and technology projects in space science of Chinese Academy of Sciences.
  The data analysis is supported in part by
  the National Key Research and Development Program of China (No. 2022YFF0503301),
  the National Natural Science Foundation of China (No. 12003074),
  {
  the Strategic Priority Program on Space Science of Chinese Academy of Sciences (No.~E02212A02S),
  the Project for Young Scientists in Basic Research of the Chinese Academy of Sciences (No.~YSBR-092),
  }
  and
  the Youth Innovation Promotion Association CAS.


\bibliographystyle{elsarticle-num}

\bibliography{main.bib}



\end{document}